\input epsf.sty
\documentstyle{mn}

\def\ltsim{\raise 2pt \hbox {$<$} \kern-1.1em \lower 4pt \hbox {$\sim$}}
\def\ltapprox{\raise 2pt \hbox {$<$} \kern-1.1em \lower 5pt \hbox {$\approx$}}
\def\gtsim{\raise 2pt \hbox {$>$} \kern-1.1em \lower 4pt \hbox {$\sim$}}
\def\gtapprox{\raise 2pt \hbox {$>$} \kern-1.1em \lower 5pt \hbox {$\approx$}}

\title[The A3528 Cluster Complex]{Radio Properties 
of the Shapley Concentration \\ IV. The A3528 Cluster Complex}

\author[T. Venturi et al.]
{T. Venturi$^{1,\star}$,  S. Bardelli$^2$, G. Zambelli$^1$,
R. Morganti$^3$, R.W. Hunstead$^4$\\
$^1$ Istituto di Radioastronomia, CNR,
via Gobetti 101, I-40129 Bologna, Italy \\
$^2$ Osservatorio Astronomico di Bologna, 
via Ranzani 1, I-40126 Bologna, Italy \\
$^3$ NFRA, Dwingeloo, Postbus 2, AA 7990, The Netherlands \\
$^4$ School of Physics, University of Sydney, NSW 2006, Australia \\
$\star$ E-mail: tventuri@ira.bo.cnr.it \\
}

\date{Received XX; Accepted XX}
\pagerange{\pageref{firstpage}--\pageref{lastpage}}
\pubyear{}

\begin{document}

\maketitle
 
\begin{abstract}

We report and discuss the results of a 22 cm radio survey carried out with the
Australia Telescope Compact Array (ATCA) covering the
A3528 complex, a chain formed by the merging ACO clusters A3528-A3530-A3532,
located in the central region of the Shapley Concentration. Simultaneous
13 cm observations are also presented. Our final catalogue includes
a total of 106 radio sources above the flux density limit of 0.8
mJy. By cross correlation with optical and spectroscopic catalogues
we found 32 optical counterparts, nine of them belonging to 
the A3528 complex. 

We explored the effects of cluster mergers on the radio emission
properties of radio galaxies by means of the radio luminosity 
function (RLF) and radio source counts. We found that the radio source
counts are consistent with the background counts, as already found
for the A3558 complex (Paper III). The RLF for this complex 
is consistent, both in shape and normalisation, with the
general cluster luminosity function for early-type galaxies
derived by Ledlow \& Owen (LO96).
This result is different from what we obtained for the A3558 merging 
complex, whose RLF is significantly lower than LO96.

We propose that the different stage of the merger is responsible
for the different RLFs in the two cluster complexes in the
core of the Shapley Concentration. The early stage of
merger for the A3528 complex, proposed by many authors, 
may have not yet affected the radio properties of
cluster galaxies, while in the more much advanced merger in the
A3558 region we actually see the effects of this process
on the radio emission.

\end{abstract}


\section{Introduction}

Cluster mergers are the most energetic and common phenomena
in the Universe, leading to an energy release of 
$\sim 10^{50-60}$ ergs on a timescale of the order of Gyrs. 
However, it is still unclear 
how this enormous energy release is dissipated.
While the effects of merging on the large scale 
properties of the hot intracluster medium (shocks, 
gradients in the temperature and changes in
the gas distribution profiles) seem to be reasonably well understood
(see for example Roettiger, Burns \& Loken 1996), a 
key question is whether it has important consequences 
also on the galaxy population properties.
Bekki (1999) suggested that merging could be important
in driving gas towards the centres of galaxies and, therefore, 
in fuelling the star formation.
Owen et al. (1999), in a comparative study of the two clusters 
A2125 and A2645, suggested that cluster merging could be 
responsible both for the high fraction of blue (i.e.
star-forming) galaxies and for the excess in the radio source 
population in A2125.

In order to explore in detail the astrophysical consequencies
of merging on galaxy radio emission,
Venturi et al. (2000, Paper III in this series)
carried out a detailed multifrequency study of the A3558 
cluster complex, at the centre of the Shapley Concentration.
This structure is formed by three Abell clusters
and two poor groups. Bardelli et al. (1998b) suggested that
it is the remnant of a cluster-cluster collision, seen just after
the first core-core encounter. This hypothesis is supported
by the dynamical state of clusters (Bardelli et al. 1998a)  
and by a filament of hot gas connecting the clusters
(Bardelli et al. 1994, Kull \& B\"ohringer 1999).
Moreover, there is marginal evidence of an excess
of blue galaxies  at the expected shock position.
Substructure analysis (Bardelli et al. 1998b) and the detailed 
dynamical study of a possible relic source (Venturi et al. 1998, Paper 
II) show that this complex is far from a relaxed state.

\noindent
Rather unexpectedly, the bivariate radio-optical
luminosity function of the A3558 complex shows a
significant deficiency of radio sources in comparison with
the cluster sample studied by 
Ledlow \& Owen 1996  (hereinafter LO96),
suggesting that merging may have ``switched-off" 
pre-existing radio sources.
These results suggest that the role of cluster mergers on
the radio properties of galaxies is not well understood yet,
and other parameters, such as, for example, the evolutionary
stage of the merger and the initial conditions (such as for example 
cluster masses) are likely to play a major role. 

Very few merging
clusters are as well studied at optical, X--ray and radio
wavelengths as done for the A3558 complex,
therefore it is of crucial importance to increase 
their number, in order to throw light on such
a complex phenomenon and better understand the relation
between cluster mergers and the radio properties
of galaxies.
With this aim in mind, we
surveyed another region of merging clusters in the core of
the Shapley Concentration, dominated by A3528.
This structure is formed by the ACO (Abell, Corwin \& Olowin 1989)
clusters A3528, A3530, A3532 and A3535. 
These clusters have been observed with ROSAT (Schindler 1996; 
Henriksen \& Jones 1996; White, Jones \& Forman 1997); moreover,
Bardelli, Zucca \& Baldi (2000)
performed a redshift survey of $\sim 700$ galaxies in order 
to determine the dynamics of the complex.

We observed the complex simultaneously at 22 cm and 13 cm
with the Australia Telescope Compact Array (ATCA). We covered the whole
region, an area of $\sim 2 \times 1$ deg$^2$ in the sky.
In this paper, the fourth of a series of papers dedicated to
this study, we present the results of our observations.
In Section 2 we summarize the global 
properties of the Abell clusters in this region; in Section 3 we
describe the observations, data reduction and 
give a catalogue of the radio sources; optical identifications
for the detected sources are also presented in this
section. 
The properties of the extended radio galaxies in the A3528 complex 
are given in Section 4; our statistical
analysis and conclusions are given in Section 5 and 6 respectively.

\par
As in previous papers in this series, 
we assume a Hubble constant H$_0$ = 100 h km s$^{-1}$Mpc$^{-1}$.
At the average redshift of the Shapley Concentration,
$\langle z \rangle =~0.05$, 
$1^{\prime\prime}$ = 0.67 kpc.

%\vskip 1 truecm \noindent

\section{The A3528 cluster complex in the Shapley Concentration} 

The A3528 cluster complex is a group of galaxy clusters located in
the core of the Shapley Concentration, northwest of the A3558 complex,
the dominant structure of the supercluster, and it is formed by the three
ACO clusters A3528, A3530, A3532.
It is at an average redshift of $\langle z \rangle = 0.0535$
($\sim$~16000 km s$^{-1}$) and has 
an elongation of $\sim$ $3^{\circ}$  in the North-South direction
($\sim$ 8 h$^{-1}$ Mpc). 
The poor cluster A3535, located northeast of A3528, although close 
in projection to the other clusters of the complex, is at a larger 
redshift, i.e. 
$\langle z \rangle ~\sim~0.07$ ($\sim$ 20000 km s$^{-1}$). 

\begin{figure}
\epsfysize=5cm
\epsfxsize=\hsize
\epsfbox{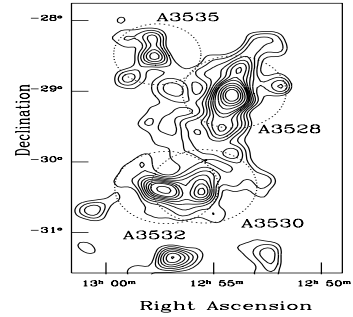}
\caption[]{Isocontours of the galaxy density to b$_J$ = 19.5 in the
A3528 complex. Dashed circles correspond to 1 Abell
radius around cluster centres. Coordinates are J2000.}
\label{fig:a3528}
\end{figure}

\noindent
Figure \ref{fig:a3528} shows the isodensity contours of the
distribution of the optical galaxies with magnitude
b$_J \le$ 19.5; the levels have been chosen in order to
highlight  the clusters. The figure shows two interesting features:
a) the distance between the centres of A3530 and A3532 is smaller
than their Abell radii, an indication at least of tidal interactions;
b) the contours of A3528 appear to be elongated in the North-South 
direction, pointing toward the A3530-A3532 system.
A substructure analysis done by Bardelli, Baldi \& Zucca (2000)
led to the conclusion that the whole chain 
is in a very complex dynamical state.
Table 1 summarizes the most important features of the
three clusters: in columns 1, 2 and 3 we
give respectively name and J2000 coordinates; 
in column 4 the Bautz-Morgan type and richness;
columns 5 and 6 show the mean heliocentric velocity
and velocity dispersion; mass, X--ray luminosity and gas
temperature are reported  respectively in columns 7, 8 and  9.

%% FIGURE 1

\medskip 

A3528 is the dominant cluster in this region, with richness
class 1 (Abell, Corwin \& Olowin 1989) and a cD galaxy with
b$_J$ = 13.6 located at its centre.
 X-ray observations (Raychaudhury et al. 1991; Schindler 1996) revealed that
the X-ray emission in A3528 is bimodal, and actually consists of two 
components, called A3528N and A3528S, whose centres
are separated by $\sim$~13~arcmin ($\sim$ 570h$^{-1}$ kpc).

It has been proposed that A3528 is an early merger.
This hypothesis is supported by the X-ray temperature gradient
in the regions facing the two structures (Schindler 1996, hereinafter S96),
and by the complex results emerging from the substructure 
analysis (Bardelli et al., 2000).

22 cm radio observations carried out by Reid, Hunstead and Pierre
(1998, hereinafter RHP98) show that five extended radio sources are 
associated with cluster galaxies. They proposed that the
head-tailed source located near the centre of A3528N was triggered
on its first orbit around the cluster centre. They therefore
concluded that the pre-merging stage of the cluster is
responsible for the nuclear activity of the central radio galaxies.

\medskip

% TABLE 1.
%\setcounter{table}{1}
\begin{table*}
\caption[]{Properties of the clusters in the A3528 chain}
\begin{flushleft}
\begin{tabular}{lllclrrrr}
\hline\noalign{\smallskip}
Cluster & RA$_{J2000}$ & DEC$_{J2000}$ & B-M Type (R) &  
$<v>$ & $\sigma_v $ & Mass & L$_X$ & T$_X$ \\
        &              &               &          &
km s$^{-1}$ & km s$^{-1}$  & 10$^{14}$h$^{-1}$M$_{\odot}$  & 
10$^{43}$ erg s$^{-1}$ & kT \\\\
%\noalign{\smallskip}
\hline\noalign{\smallskip}
A3528   & 12 54 34     & $-$29 08 30   &   II (1)  & 
16332$^{+72}_{-116}$ & 955$^{+86}_{-86}$ & N: 3.1$^a$ & N: 3.2$^a$ 
& N: 2.7$^a$ \\
        &              &               &          &    
        &               & S: 3.9$^a$ & S: 4.1$^a$ & S: 2.9$^a$ \\
A3530   & 12 55 31     & $-$30 19 53   &  I-II (0)   & 
16274$^{+78}_{-110}$ & 730$^{+143}_{-53}$ & 1.1$^b$  &    2.4$^c$ & 3.2$^c$ \\
A3532   & 12 57 22     & $-$30 22 03   & II-III (0)  & 
16605$^{+76}_{-161}$ & 447$^{+42}_{-22}$ & 1.7$^b$  &    7.2$^c$ & 4.4$^c$ \\
\noalign{\smallskip}
\hline
\end{tabular}

Notes to Table 1.

$<v>$ and $\sigma_v$ for all clusters are taken from Bardelli et al. (2000).

$^a$ Henriksen \& Jones (1996); 

$^b$ Ettori, Fabian \& White (1997)

$^c$ White, Jones \& Forman (1997).

\end{flushleft}
\end{table*}
%------- end of Table 1

A3530 and A3532 are aligned approximately east-west,
perpendicular to the projected extension of A3528. They 
are both richness class 0 clusters.
A3530 is dominated by two very bright elliptical galaxies, each 
surrounded by an extended envelope, and probably interacting,
while a dumbell galaxy is located at the centre of A3532.

\noindent
Both clusters are characterised by X-ray emission, which
peaks at the optical density peak. 
No radio observations exist in the literature. 
However, the dumbell galaxy located at the centre of A3532 is
a well known radio source with extended morphology
(Gregorini et al. 1994).

\medskip

\section{The A3528 radio survey}

\subsection{22 and 13 cm  observations and data reduction}

% FIGURE 2
\begin{figure}
\epsfysize=5cm
\epsfxsize=\hsize
\epsfbox{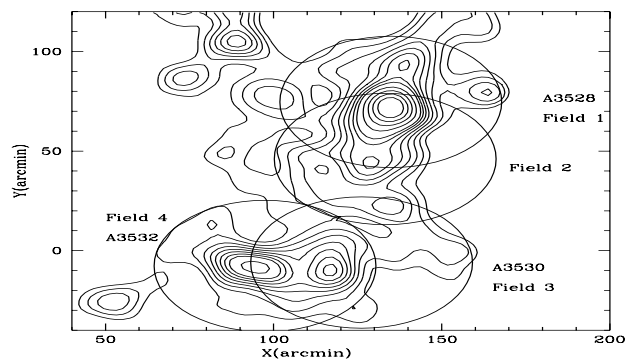}
\caption[]{Pointings of our survey (labelled as in Table 2)
superimposed on the isodensities
of the A3528 cluster complex. The radius of the circles is 33 arcmin,
corresponding to the ATCA primary beam at 22 cm.}
\label{fig:point}
\end{figure}

The observations of the A3528 complex at 22 and 13 cm
were carried out with the
Australia Telescope Compact Array (ATCA). At 22 cm (1.38 GHz)
the region was well covered with 4 pointings, characterised by different 
array configurations and exposure times. The details of the observations
are summarized in Table 2, while the spatial coverage is shown
in Figure \ref{fig:point}, overlaid on the optical isodensities.
Note that the overlapping regions between adjacent 
fields ensure an almost perfect coverage of the whole cluster chain.

% TABLE 2.
%\setcounter{table}{2}
\begin{table*}
\caption[]{Details of the observations}
\begin{flushleft}
\begin{tabular}{lllcrccc}
\hline\noalign{\smallskip}
Field & RA$_{J2000}$ & DEC$_{J2000}$ & Array    & Min. Max. Baseline & Time & 
rms (22 cm)  & rms (13 cm) \\
\#    &              &               &          &    m          &   hr & 
mJy/beam     & mJy/beam    \\
\noalign{\smallskip}
\hline\noalign{\smallskip}
   1   &   12 54 24  &  $-$29 01 19  &   6C     & 153 - 6000    &   2  &  
0.10   & 0.20  \\
   2   &   12 54 30  &  $-$29 29 59  &1.5B + 6C &  31 - 6000  & $4\times4$ &  
0.17   & 0.12  \\
   3   &   12 55 00  &  $-$30 22 00  &1.5B + 6C &  31 - 6000  & $4\times4$ &  
0.10   & 0.10  \\
   4   &   12 57 13  &  $-$30 23 44  &   6C     & 153 - 6000  &     2      &
0.10   & 0.14  \\
\noalign{\smallskip}
\hline
\end{tabular}
\end{flushleft}
\end{table*}
%------- end of table 2

Fields \#1 and \#4 are archive ATCA data and refer to observations
carried out in March 1994 (see RHP98) which we re-analysed starting
from the original uncalibrated u-v data.
We observed fields \#2 and \#3 in May and July 1995 
with the ATCA in two different array
configurations (see Table 2) which we combined in order to
improve the u-v coverage and sensitivity.
The resolution in each field is $\sim 10^{\prime\prime} 
\times 6^{\prime\prime}$. However,
thanks to the better u-v coverage at short spacings,
the sensitivity to extended
structure is higher in fields \#2 and \#3.
%In particular, the largest detectable structure in these
%fields is ... arcsec, to be compared to 40 arcsec in fields
%\#1 and \#4

\medskip

Each observation was carried out with a 128 MHz bandwidth,
using a correlator configuration with 32 4-MHz wide channels 
to minimise bandwidth smearing at large distances
from the field centre. B1934$-$638 was used as primary flux calibrator,
with an assumed flux density S$_{22cm}$ = 14.9 Jy.
The data reduction was carried out with the MIRIAD package
(Sault, Teuben \& Wright 1995),
which is particularly suited for the ATCA observations.
The image analysis was carried out with the
AIPS package.
The average noise in the final images is $\sim$ 0.1 mJy/beam 
in all fields except field \#2, where the presence of two strong
sources at the field edge increases the rms to
0.17 mJy/beam.

We consider as reliable all detections with peak
flux density S$_{22cm} \ge$ 0.8 mJy/beam,
which corresponds to $\sim 5\sigma$ confidence level in field \#2.

In order to compensate for the sensitivity loss towards the
field edge due to the primary beam attenuation, we corrected
the flux densities using the analytical formula
of the ATCA primary beam attenuation given by
Wieringa \& Kesteven (1992). 

\medskip

The uncertainty in the flux density measurement $\Delta S$ is a
function of the map noise, the source flux density and
the residual calibration error. We have assumed that

$$\Delta S = \sqrt{a^2+(bS)^2}$$

\noindent
where $a$ is the map noise, $b$ is the residual calibration error,
estimated to be of the order of 1\%, and $S$ is the source
flux density.
We point out however that 
a detailed analysis on the internal accuracy 
of the flux density measures for the sources in the
ATESP survey (Prandoni et al. 2000a), carried out with the
ATCA at 1.4 GHz at resolution comparable with the data
presented in this paper, showed that the uncertainties
in the flux density measures obtained from Gaussian fits 
for sources with low signal to noise ratio, i.e. SNR $\le 10$,
can be as high as 10$~-~$15 \%.

The radio position error depends on the beam size and the
source flux density. With the parameters of our observations
the position uncertainty for the weakest sources in the sample is 
$\Delta\alpha \sim 0.5^{\prime\prime}$ and
$\Delta\delta \sim 1^{\prime\prime}$ in right ascension and
declination respectively.

\medskip
Simultaneous 13 cm (2.38 GHz) observations were carried out 
in each configuration, and the data reduction and image analysis
were carried out as for the 22 cm dataset. 
The adopted flux density for the primary calibrator B1934$-$638
is S$_{13cm}$ = 11.6 Jy. The field of view of
the ATCA at 13 cm is 22$^{\prime}$, so only the central
part of the 22 cm fields was covered by the 13 cm observations.
The 13 cm rms in each field is reported in Table 2.
The resolution of our 13 cm images is 
$\sim 6.0^{\prime\prime} \times 3.5^{\prime\prime}$. On the basis
of the rms reported in Table 2 we placed
a detection limit of S$_{13cm}$ = 0.8 mJy.

\subsection{The sample of radio sources at 22 cm}

We detected a total of 106 radio sources at 22 cm above
the flux density limit of 0.8 mJy.
We point out that the 22 cm mJy survey presented here and
in Paper III of this series is the deepest wide area surveys available
in a cluster of galaxies type environment (see Prandoni et al. 2000b
for a review of the mJy and sub-mJy surveys available
at this frequency).

The source list is reported in Table 3, where we give name and position
(columns 1, 2 and 3), integrated flux density at 22 cm and 13 cm 
corrected for the primary beam attenuation (columns 4 and 5), 
and the radio morphology in column 6. 
From our 22 cm images we classified the radio
morphology of the sources in our sample as follows: unresolved,
resolved, D=double; HT=head tail; WAT=wide angle tail.
For the double sources in the sample we give the position of
the radio barycentre, and for the remaining extended sources we
give the position of the radio peak.

The 13 cm flux density measurements are available only for a 
fraction of sources in our list, i.e. all sources with a distance
smaller than $\sim 20$ arcmin from the respective field centre.
As clear from Table 3, another fraction of the 22 cm sources surveyed
at 13 cm has a flux density S$_{13cm}$ below the sensitivity limit
of our 13 cm observations. Finally a few sources have S$_{13cm} <$ 0.8
mJy/beam (before the primary beam correction), but they are 
undoubtely real. We marked these flux densities with $\star$.

Most of the sources detected in the 22 cm  survey are unresolved;
there are  18 extended radio sources, corresponding
to $\sim 17$ \% of the total.
Among these extended radio sources we identify three
wide-angle tail (WAT) sources, three head-tails (HT) and 
five doubles. 
The remaining extended sources have more complex morphologies,
and require observations at other frequencies and resolutions
for a clear morphological classification.

We note that  the flux densities reported in Table 3 for 12 sources
were also given in RHP98. The values we derived for
those sources after re-analysis of the archive data set
are a few percent higher than those already 
published. For all sources in our sample we also compared our
flux density measurements with those obtained from the 
NRAO VLA Sky Survey (NVSS) images (Condon et al. 1998), 
and found that the ratio $S_{ATCA} \over S_{NVSS}$ 
is in the range 0.85$~-~$1, depending on the source structure,
consistent with the lower resolution and different u-v coverage
of NVSS. Given the better agreement between our values and
those obtained from NVSS and 
our more homogeneous analysis of the radio data, 
we now prefer the flux densities reported in Table 3
for all sources in the A3528 complex.

%
%
% BEGIN TABLE 3
%\setcounter{table}{3}
\begin{table*}
\caption[]{Source list and flux density values}
\begin{flushleft}
\begin{tabular}{lllrrr}
\hline\noalign{\smallskip}
Name   & RA$_{J2000}$  & DEC$_{J2000}$ & S$_{22 cm}$ & S$_{13 cm}$ & R Morph \\
       &               &               &     mJy     &    mJy      &         \\
\\
\noalign{\smallskip}
\hline\noalign{\smallskip}
J1252$-$2927  & 12 52 04.5  & $-$29 27 54  &   77.7  &    -    &  unres.\\
J1252$-$2928a & 12 52 04.7  & $-$29 28 24  &  149.6  &    -    &  unres.\\
J1252$-$2902  & 12 52 17.9  & $-$29 02 10  &   10.1  &    -    &  unres.\\ 
J1252$-$2908  & 12 52 33.2  & $-$29 08 05  &  101.1  & $<$ 0.8 &   D    \\
J1252$-$2917  & 12 52 48.3  & $-$29 17 01  &   10.3  & $<$ 0.8 &  unres.\\
J1252$-$2928b & 12 52 56.7  & $-$29 28 00  &   12.6  & $<$ 0.8 &  unres.\\ 
J1253$-$2902  & 12 53 05.4  & $-$29 02 36  &    6.0  & $<$ 0.8 &  unres.\\
J1253$-$3018a & 12 53 06.6  & $-$30 18 45  &  141.7  &    -    &  unres.\\
J1253$-$2853a & 12 53 10.7  & $-$28 53 26  &   35.4  &   18.3  &  unres.\\
J1253$-$3018b & 12 53 15.4  & $-$30 18 07  &   16.6  &    -    &  unres.\\
J1253$-$2836  & 12 53 16.0  & $-$28 36 47  &    6.5  &    -    &  unres.\\
J1253$-$2850  & 12 53 18.1  & $-$28 50 20  &    4.2  & $<$ 0.8 &  unres.\\
J1253$-$3007  & 12 53 19.2  & $-$30 07 23  &    6.6  &    -    &  unres.\\
J1253$-$2859  & 12 53 19.5  & $-$28 59 52  &   78.4  &   82.2  &  unres.\\
J1253$-$2937  & 12 53 25.6  & $-$29 37 00  &   29.2  &   19.5  &  unres.\\
J1253$-$2921  & 12 53 28.7  & $-$29 21 18  &   30.1  &   15.2  &     D  \\
J1253$-$2936  & 12 53 29.6  & $-$29 36 49  &    6.9  &    4.0  &  unres.\\
J1253$-$2938  & 12 53 32.3  & $-$29 38 30  &   36.4  &   25.0  &  unres.\\
J1253$-$3025  & 12 53 40.5  & $-$30 25 33  &   29.6  &    -    &  unres.\\
J1253$-$2900  & 12 53 44.3  & $-$29 00 00  &    1.5  &    2.0  &  unres.\\
J1253$-$2930  & 12 53 46.5  & $-$29 30 10  &    2.1  &    1.9  &  unres.\\
J1253$-$3010  & 12 53 47.1  & $-$30 10 07  &   11.7  & $<$ 0.8 &  unres.\\
J1253$-$2854  & 12 53 47.3  & $-$28 54 20  &   18.3  &   20.4  &  unres.\\
J1253$-$3013  & 12 53 48.6  & $-$30 13 55  &    2.3  & $<$ 0.8 &  unres.\\ 
J1253$-$2841  & 12 53 49.1  & $-$28 41 57  &    3.0  & $<$ 0.8 &  unres.\\
J1253$-$3012a & 12 53 49.6  & $-$30 12 38  &   16.2  & $<$ 0.8 &  unres.\\
J1253$-$2853b & 12 53 49.7  & $-$28 53 30  &    4.4  &    5.4  &  unres.\\
J1253$-$3012b & 12 53 50.1  & $-$30 12 57  &   12.8  & $<$ 0.8 &  unres.\\
J1253$-$2848  & 12 53 54.8  & $-$28 48 54  &    1.4  & $<$ 0.8 &  unres.\\
J1253$-$3017  & 12 53 57.7  & $-$30 17 51  &    3.5  &2.4$\star$& unres.\\
J1253$-$2935  & 12 53 58.6  & $-$29 35 33  &   23.4  &   17.8  &  unres.\\
J1254$-$2858a & 12 54 01.8  & $-$28 58 13  &    3.9  &    3.2  &  unres.\\
J1254$-$2927  & 12 54 03.2  & $-$29 27 22  &   59.3  &   44.2  &    res.\\  
J1254$-$2858b & 12 54 05.7  & $-$28 58 03  &    3.2  & $<$ 0.8 &  unres.\\
J1254$-$3049  & 12 54 20.3  & $-$30 49 26  &   45.4  &     -   &  unres.\\
J1254$-$3025  & 12 54 20.8  & $-$30 25 10  &    2.0  &1.1$\star$& unres.\\
J1254$-$2904  & 12 54 21.2  & $-$29 04 16  &  295.9  &  142.8  &    HT  \\
J1254$-$3011  & 12 54 21.4  & $-$30 11 53  &    2.6  &2.2$\star$& unres.\\  
J1254$-$2900  & 12 54 22.1  & $-$29 00 48  &   230.9 &  142.4  &      D \\
J1254$-$2901a & 12 54 22.9  & $-$29 01 02  &   110.5 &   57.9  &  HT \\
J1254$-$2859  & 12 54 23.2  & $-$28 59 39  &     8.7 &    7.2  & unres.\\
J1254$-$2933  & 12 54 37.2  & $-$29 33 32  &     3.1 &    1.8  & unres.\\
J1254$-$3012  & 12 54 38.2  & $-$30 12 46  &     2.0 &    1.6  & unres.\\
J1254$-$3015  & 12 54 38.7  & $-$30 15 13  &     1.1 &    1.2  & unres.\\
J1254$-$3045a & 12 54 40.1  & $-$30 45 54  &     7.9 &    -    & unres.\\
J1254$-$2901b & 12 54 40.2  & $-$29 01 46  &    14.2 & $<$ 0.8 &   res.\\ 
J1254$-$2939  & 12 54 40.6  & $-$29 39 15  &     5.4 &    3.1  & unres.\\
J1254$-$2913  & 12 54 41.0  & $-$29 13 39  &   936.7 &  538.4  &   WAT \\
J1254$-$3019  & 12 54 42.6  & $-$30 19 28  &     5.7 &    3.5  & unres.\\
J1254$-$3045b & 12 54 42.9  & $-$30 45 23  &    15.3 &    -    & unres.\\
J1254$-$3008  & 12 54 43.8  & $-$30 08 39  &     1.6 & $<$ 0.8 & unres.\\
J1254$-$3028  & 12 54 46.7  & $-$30 28 35  &     2.7 &    1.8  & unres.\\
J1254$-$2949  & 12 54 50.4  & $-$29 49 24  &   292.6 &  153.0  &   res.\\
J1254$-$3042  & 12 54 50.4  & $-$30 42 11  &   122.6 &   63.1  &   WAT \\ 
J1254$-$2916  & 12 54 51.4  & $-$29 16 20  &    62.1 &    49.6 &    HT \\
J1254$-$3046  & 12 54 51.6  & $-$30 46 53  &    60.0 & $<$ 0.8 & unres.\\
\noalign{\smallskip}	      		     	       
\hline
\end{tabular}
\end{flushleft}
\end{table*}
%
%
%
%-- TABLE 3 - continued
\setcounter{table}{2}
\begin{table*}
\caption[]{ Continued}
\begin{flushleft}
\begin{tabular}{lllrrr}
\hline\noalign{\smallskip}
Name  & RA$_{J2000}$  & DEC$_{J2000}$ & S$_{22 cm}$ & S$_{13 cm}$ & R Morph \\
      &               &               &  mJy        &             &         \\
\\
\noalign{\smallskip}
\hline\noalign{\smallskip}

J1255$-$2907  & 12 55 00.7  & $-$29 07 13  &    2.7 &    1.9  &  unres.\\
J1255$-$3024  & 12 55 03.9  & $-$30 24 33  &    3.0 &    2.1  &  unres.\\ 
J1255$-$3019  & 12 55 06.1  & $-$30 19 05  &    2.2 &    1.6  &  unres.\\
J1255$-$3018  & 12 55 06.8  & $-$30 18 11  &    2.0 &    1.3  &  unres.\\
J1255$-$3007  & 12 55 09.1  & $-$30 07 53  &    2.4 & $<$ 0.8 &  unres.\\
J1255$-$2934  & 12 55 09.9  & $-$29 34 06  &   60.1 &   36.6  &   D    \\
J1255$-$3023  & 12 55 13.2  & $-$30 23 53  &    1.2 &    0.9  &  unres.\\
J1255$-$2908  & 12 55 14.2  & $-$29 08 19  &    3.2 &    2.2  &  unres.\\
J1255$-$2924  & 12 55 14.9  & $-$29 24 43  &    2.8 &    2.0  &  unres.\\
J1255$-$2933  & 12 55 16.4  & $-$29 33 32  &    2.3 &    2.0  &  unres.\\
J1255$-$2919  & 12 55 21.1  & $-$29 19 53  &    5.3 & $<$ 0.8 &  unres.\\
J1255$-$3034  & 12 55 22.1  & $-$30 34 45  &    2.1 & $<$ 0.8 &  unres.\\
J1255$-$3041  & 12 55 24.2  & $-$30 41 15  &   12.8 &    -    &  unres.\\
J1255$-$3005  & 12 55 28.1  & $-$30 05 53  &   35.0 &   19.9  &  unres.\\
J1255$-$2939  & 12 55 30.9  & $-$29 39 25  &    5.7 &    6.7  &  unres.\\
J1255$-$3022  & 12 55 34.9  & $-$30 22 17  &    1.8 &    1.0  &  unres.\\
J1255$-$2923  & 12 55 35.4  & $-$29 23 43  &   14.6 &   12.4  &  unres.\\
J1255$-$3040a & 12 55 35.6  & $-$30 40 31  &    9.6 &    -    &  unres.\\ 
J1255$-$2855  & 12 55 37.8  & $-$28 55 56  &   31.2 &   15.5  &  unres.\\
J1255$-$2919  & 12 55 44.8  & $-$29 19 58  &   20.8 &   26.7  &  unres.\\
J1255$-$2943  & 12 55 46.8  & $-$29 43 06  &   40.2 &    -    &  unres.\\
J1255$-$3042  & 12 55 47.3  & $-$30 42 59  &   41.4 &    -    &  unres.\\
J1255$-$3012  & 12 55 53.6  & $-$30 12 45  &   74.9 &   37.4  &     D  \\
J1255$-$3040b & 12 55 58.0  & $-$30 40 19  &    6.7 &    -    &  unres.\\
J1255$-$3008  & 12 55 59.8  & $-$30 08 51  &    3.1 & $<$ 0.8 &  unres.\\
J1256$-$3009  & 12 56 10.6  & $-$30 09 57  &   27.5 & see text&    D?  \\
J1256$-$2909a & 12 56 10.9  & $-$29 09 57  &   10.8 &    -    &   unres.\\
J1256$-$2851a & 12 56 15.7  & $-$28 51 03  &   13.8 &    -    &   unres.\\
J1256$-$2851b & 12 56 17.0  & $-$28 51 46  &   81.4 &    -    &     res.\\
J1256$-$2852  & 12 56 17.8  & $-$28 52 15  &   17.1 &    -    &   unres.\\
J1256$-$3014  & 12 56 19.2  & $-$30 14 39  &    3.6 &    -    &   unres.\\
J1256$-$2909b & 12 56 24.8  & $-$29 09 40  &   23.2 &    -    &   unres.\\
J1256$-$3039  & 12 56 27.3  & $-$30 39 05  &    3.7 &    -    &   unres.\\
J1256$-$2912  & 12 56 28.3  & $-$29 12 22  &   19.5 &    -    &   unres.\\
J1256$-$2926  & 12 56 38.9  & $-$29 26 40  &   31.5 &    -    &   unres.\\
J1256$-$3040  & 12 56 48.3  & $-$30 40 47  &    7.1 & $<$ 0.8 &   unres.\\ 
J1257$-$3022  & 12 57 04.3  & $-$30 22 29  &   17.0 &    9.3  &   unres.\\
J1257$-$3026  & 12 57 08.2  & $-$30 26 17  &   10.1 &   16.3  &   unres.\\
J1257$-$3012  & 12 57 08.3  & $-$30 12 05  &    2.1 & $<$ 0.8 &   unres.\\
J1257$-$3047  & 12 57 13.6  & $-$30 47 46  &   70.0 &    -    &   unres.\\
J1257$-$3020  & 12 57 19.6  & $-$30 20 47  &    3.6 & $<$ 0.8 &   unres.\\
J1257$-$3021  & 12 57 22.5  & $-$30 21 45  & 1056.5 &  651.7  &   WAT   \\
J1257$-$3013  & 12 57 28.6  & $-$30 13 07  &   23.6 &    2.6  &     res.\\
J1257$-$3011  & 12 57 33.5  & $-$30 11 19  &    3.2 & $<$ 0.8 &   unres.\\ 
J1257$-$3021  & 12 57 36.3  & $-$30 21 45  &    1.3 & $<$ 0.8 &   unres.\\
J1257$-$3030  & 12 57 53.9  & $-$30 30 11  &   12.1 &    3.6  &   unres.\\
J1258$-$3034  & 12 58 19.5  & $-$30 34 30  &   46.9 &   49.9  &   unres.\\
J1258$-$3027  & 12 58 26.2  & $-$30 27 56  &   44.3 &   25.0  &     res.\\
J1258$-$3012  & 12 58 36.4  & $-$30 12 05  &    9.2 &    -    &    unres.\\
J1258$-$3016  & 12 58 57.3  & $-$30 16 19  &   12.7 &    -    &    unres.\\
\noalign{\smallskip}					  
\hline							   
\end{tabular}						   
\end{flushleft}
\end{table*}

% end of table 3

\subsection{Optical identifications}

We carried out the optical identifications of the radio sources	       
in our 22 cm sample by cross-correlating the source positions
with galaxies listed in the  COSMOS/UKST
Southern Sky Object Catalogue (Yentis et al 1992), which
contains objects down to b$_J$ = 21
and has a claimed positional accuracy of $\sim$ 0.25 arcsec.
Given that the uncertainty due to the transposition of the sky
image on the plate frame may lead to larger errors, for our study we 
assume a mean optical positional
error of 1.5 arcsec.
Since COSMOS is incomplete at faint magnitudes
and could miss bright galaxies, all the radio sources were 
overplotted on the  Digitized Sky Survey
(DSS) and scrutinized by eye, so as to make sure no faint
identification was missed.

\noindent
Given the uncertainties in the radio and optical positions
we adopt the parameter $R$, defined as:

$$ R^2 = {{\Delta^2_{r-o}} \over {\sigma^2_g + \sigma^2_r}}$$

\noindent
to assess the reliability of the optical identifications.
In the above formula, $\Delta_{r-o}$ is the positional
offset, $\sigma_g$ is the galaxy position error and
$\sigma_r$ is the uncertainty in the radio position.
For point-like radio sources we consider as reliable
identifications all matches with $R \le 3$.
The list of the radio-optical identifications is given
in Table 4, where we report the radio and optical name (column 1), J2000 radio
and optical coordinates (columns 2 and 3), radio flux densities and
b$_J$ mag (column 4), 22 cm radio power logP$_{22cm}$ and 
absolute magnitude B$_J$ (column 5),
radio morphology and optical type (column 6),
$R$ and the radial velocity (column 7).

For five extended radio sources we found $R > 3$ (see notes to 
Table 5 and Section 4.1).
However, we consider them reliable identifications, given the
extension of the radio emission and the large extent of the underlying 
galaxy.

\medskip

\begin{figure}
\epsfysize=5cm
\epsfxsize=\hsize
\epsfbox{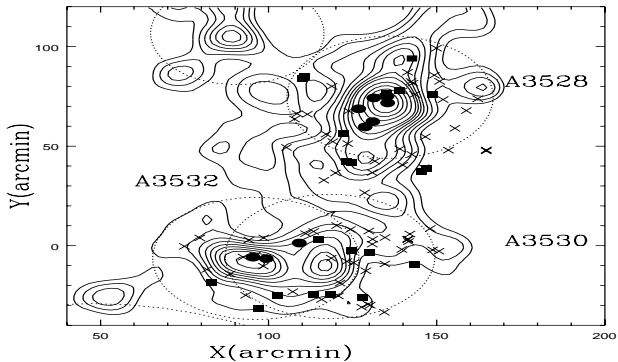}
\caption[]{Radio sources detected in the A3528 region.
Filled circles represent radio galaxies with measured redshift,
squares refer to those with unknown redshift and crosses
to radio sources without optical counterpart. Dotted circles are 
the Abell radius around each cluster centre.}
\label{fig:sources}
\end{figure}

\begin{figure}
\epsfysize=5cm
\epsfxsize=\hsize
\epsfbox{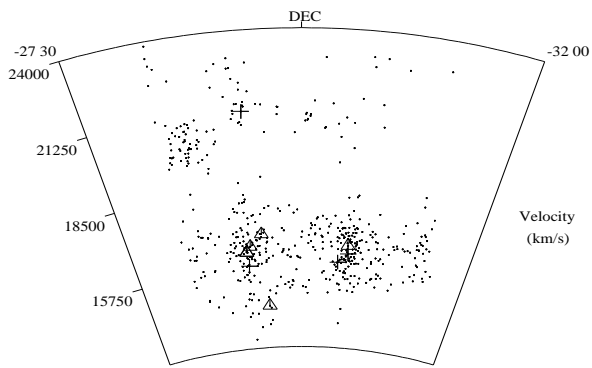}
\caption[]{Distribution of A3528 radio sources in 
velocity space. Dots are the optical galaxies; triangles are the 
extended galaxies; crosses are the point-like radio
galaxies.}
\label{fig:wedge}
\end{figure}

We found 32 optical counterparts i.e. $\sim$ 30\% of our sample.
Nine radio galaxies are located in the Shapley Concentration,
one is a background galaxy aligned with A3528 but located in A3532, 
and the distance of  the remaining 22 counterparts is unknown.
Figure \ref{fig:sources} shows the distribution of the radio sources
in the A3528 complex while in Figure \ref{fig:wedge} we report 
the distribution of the A3528 optical galaxies in velocity space 
(versus declination).

% FIGURE 3

% BEGIN TABLE 4
%\setcounter{table}{4}
\begin{table*}
\caption[]{Optical Identifications}
\begin{flushleft}
\begin{tabular}{lllrrrccc}
\hline\noalign{\smallskip}
Radio Name & RA$_{J2000}$ & DEC$_{J2000}$ & S$_{22 cm}$ & logP$_{22 cm}$ & Radio Type & $R$ \\
           &              &               &   mJy      & W Hz$^{-1}$ &  & \\
Opt.  Name & RA$_{J2000}$ & DEC$_{J2000}$ & $b_J$ &  B$_J$  & Opt.  Type & v (km s$^{-1})$\\
 \\
\noalign{\smallskip}
\hline\noalign{\smallskip}
J1253$-$2859  & 12 53 19.5 & $-$28 59 52 &   78.4 &      &       & 0.89  \\
              & 12 53 19.4 & $-$28 59 53 &  19.64 &      &       &       \\
              &            &               &        &      &       &       \\
J1253$-$2937  & 12 53 25.6 & $-$29 37 00 &   29.2 &      &       & 0.91  \\
              & 12 53 25.7 & $-$29 37 01 &  19.21 &      &       &       \\
              &            &               &        &      &       &       \\
J1253$-$2938  & 12 53 32.3 & $-$29 38 30 &   36.4 &      &       & 1.33  \\
              & 12 53 32.3 & $-$29 38 33 &  19.87 &      &       &       \\
              &            &               &        &      &       &       \\
J1253$-$3025  & 12 53 40.5 & $-$30 25 33 &   29.6 &      &       & 0.88  \\
              & 12 53 40.4 & $-$30 25 34 &  20.38 &      &       &       \\
              &            &               &        &      &       &       \\
J1253$-$2841  & 12 53 49.1 & $-$28 41 57 &   3.0  &      &       & 1.47  \\
              & 12 53 48.9 & $-$28 41 57 &  19.34 &      &       &       \\
              &            &               &        &      &       &       \\
J1254$-$2858b & 12 54 05.7 & $-$28 58 03 &   3.2  &      &       & 2.39  \\
              & 12 54 05.6 & $-$28 58 07 &  17.33 &      &       &       \\
              &            &               &        &      &       &       \\
J1254$-$2904 & 12 54 21.2 & $-$29 04 16 & 295.9 & 23.96& Ext &7.88 $^{(a)}$  \\
\#2063 & 12 54 20.3 & $-$29 04 08 &  16.88 &$-$19.21&     &16521  \\
              &            &               &        &      &       &       \\
J1254$-$2900 & 12 54 22.1 & $-$29 00 48 &230.9 & 23.84&  D  & 0.67 $^{(b)}$ \\
\#2209 & 12 54 22.1 & $-$29 00 47 &  14.61 &$-$21.46&  db &16336  \\
              &            &               &        &      &       &       \\
J1254$-$2901a & 12 54 22.9 & $-$29 01 02 & 110.5 & 23.52& HT & 1.39 $^{(c)}$ \\
     & 12 54 22.9 & $-$29 01 00   &   16.5 &$-$19.56&  db &16319   \\
              &            &               &        &      &       &       \\
J1254$-$2859  & 12 54 23.2 & $-$28 59 39 &   8.7  &      &       & 0.89  \\
              & 12 54 23.2 & $-$28 59 41 &  20.23 &      &       &       \\
              &            &               &        &      &       &       \\
              &            &               &        &      &       &       \\
J1254$-$2901b & 12 54 40.2 & $-$29 01 46 & 14.2  & 22.62&   & 3.03 $^{(d)}$  \\
\#2354        & 12 54 40.6 & $-$29 01 48 & 17.29  &$-$18.71&     &15849  \\
              &            &               &        &      &       &       \\
J1254$-$2913 & 12 54 41.0 & $-$29 13 39 & 936.7 & 24.48& WAT & 0.33 $^{(e)}$ \\
\#2370 & 12 54 41.0 & $-$29 13 39 &  14.32 &$-$21.82&     &16923  \\
              &            &               &        &      &       &       \\
J1254$-$3019  & 12 54 42.6 & $-$30 19 28 &   5.7  &      & Ris   & 1.33  \\
              & 12 54 42.6 & $-$30 19 31 & 14.08  &      & S0?   &       \\
              &            &               &        &      &       &       \\
J1254$-$3042 & 12 54 50.4 & $-$30 42 11 &  122.6  &       & WAT? & $^{(f)}$ \\
             & 12 54 50.1 & $-$30 42 14 &   23.04 &       &       & 2.60  \\
             & 12 54 50.3 & $-$30 42 12 &   20.12 &       &       & 0.45  \\
      	     &         	  &   	          &         &       &       &       \\
J1254$-$2916 & 12 54 51.4 & $-$29 16 20 & 62.1 & 23.17 &  HT & 7.61 $^{(g)}$ \\
\#2616& 12 54 52.4 & $-$29 16 16 &  15.82  &$-$19.98&      &14443  \\
      	     &         	  &   	          &         &       &       &       \\
J1255$-$2907 & 12 55 00.7 & $-$29 07 13 &    2.7  & 22.14 &       & 1.23  \\
\#2607       & 12 55 00.6 & $-$29 07 15 &  18.49  &$-$18.14&      &21193  \\
      	     &         	  &   	          &         &        &      &       \\
J1255$-$3018 & 12 55 06.8 & $-$30 18 11 &   2.0   &        &      & 1.06  \\
             & 12 55 06.7 & $-$30 18 12 &  21.65  &        &      &       \\
      	     &         	  &   	          &         &        &      &       \\
J1255$-$2934 & 12 55 09.9 & $-$29 34 06 &   60.1  &      &  D   & $^{(h)}$   \\
    & 12 55 10.6 & $-$29 34 06 &  21.87  &        &  Q?  & 5.10  \\
             & 12 55 09.3 & $-$29 34 06 &  19.61  &        &  Q?  & 4.37  \\
      	     &         	  &   	          &         &        &      &       \\

\noalign{\smallskip}
\hline
\end{tabular}

\end{flushleft}
\end{table*}
%------ continuazione tab. 4
\setcounter{table}{3}
\begin{table*}
\caption[]{Optical Identifications. Continued}
\begin{flushleft}
\begin{tabular}{lllrrrccc}
\hline\noalign{\smallskip}
Radio Name & RA$_{J2000}$ & DEC$_{J2000}$ & S$_{22 cm}$ & logP$_{22 cm}$ & Radio Type & $R$ \\
           &              &               &   mJy      & W Hz$^{-1}$ &  & \\
Opt.  Name & RA$_{J2000}$ & DEC$_{J2000}$ & $b_J$ &  B$_J$  & Opt.  Type & v (km s$^{-1})$\\
 \\
\noalign{\smallskip}
\hline\noalign{\smallskip}
J1255$-$2933 & 12 55 16.4 & $-$29 33 32 &   2.3   &        &      & 2.09  \\
             & 12 55 16.2 & $-$29 33 35 &  19.78  &        &      &       \\
      	     &         	  &   	          &         &        &      &       \\
J1255$-$2919 & 12 55 21.1 & $-$29 19 53 &   5.3   &        &      & 2.71  \\
             & 12 55 21.4 & $-$29 19 56 &  20.04  &        &      &       \\
 	     &         	  &               &         &        &      &       \\
J1255$-$3040a& 12 55 35.6 & $-$30 40 31 &   9.6   &        &      & 0.79  \\
             & 12 55 35.5 & $-$30 40 32 &  21.62  &        &      &       \\
      	     &         	  &   	          &         &        &      &       \\
J1255$-$3012 & 12 55 53.6 & $-$30 12 45 &  74.9   &        &      & 1.48  \\
             & 12 55 53.4 & $-$30 12 44 &  20.00  &        &      &       \\
      	     &         	  &   	          &         &        &      &       \\
J1255$-$3040b& 12 55 58.0 & $-$30 40 19 &   6.7   &        &      & 2.23  \\
             & 12 55 58.1 & $-$30 40 15 &  22.63  &        &      &       \\
      	     &         	  &   	          &         &        &      &       \\
J1256$-$2851a& 12 56 15.7 & $-$28 51 03 &  13.8   &        &      & 0.79  \\
             & 12 56 15.8 & $-$28 51 03 &  20.41  &        &      &	    \\
      	     &         	  &   	          &         &        &      &       \\
J1256$-$2851b& 12 56 17.0 & $-$28 51 46 &  81.4   &        &      & 2.18  \\
             & 12 56 17.2 & $-$28 51 49 &  20.58  &        &      &	    \\
      	     &         	  &               &  	    &        &      &       \\
J1256$-$2852 & 12 56 17.8 & $-$28 52 15 &   17.1  &        &      & 1.94  \\
             & 12 56 17.8 & $-$28 52 11 &  20.71  &        &      &       \\
      	     &         	  &   	          &         &        &      &       \\
J1256$-$3014 & 12 56 19.2 & $-$30 14 39 &   3.6   &  22.02 &      & 1.00  \\
\#3577       & 12 56 19.2 & $-$30 14 41 &  17.19  &$-$18.82&      &15950  \\
      	     &         	  &   	          &         &        &      &       \\
J1256$-$3040 & 12 56 48.3 & $-$30 40 47 &   7.1  &        &       & 1.32  \\
             & 12 56 48.4 & $-$30 40 49 & 21.89  &        &       &       \\
      	     &         	  &   	          &        &        &       &       \\
J1257$-$3022 & 12 57 04.3 & $-$30 22 29 & 17.0   & 22.71  &       & 0.05  \\
\#4182       & 12 57 04.3 & $-$30 22 29 & 16.48  &$-$19.57&       &16251  \\
      	     &         	  &   	          &        &        &       &       \\
J1257$-$3047 & 12 57 13.6 & $-$30 47 46 & 70.0   &        &       & 2.34  \\
             & 12 57 13.3 & $-$30 47 47 & 18.89  &        &       &       \\
      	     &         	  &   	          &        &        &       &       \\
J1257$-$3021 & 12 57 22.5 & $-$30 21 45 &1056.5 &24.51  & WAT& 8.90 $^{(i)}$ \\
\#4327& 12 57 21.3 & $-$30 21 48 &  14.49 &$-$21.60&  db   &16540  \\
      	     &         	  &   	          &        &        &       &       \\
J1258$-$3034 & 12 58 19.5 & $-$30 34 30 & 46.9   &        &       & 1.19  \\
             & 12 58 19.4 & $-$30 34 31 & 18.17  &        &       &       \\

\noalign{\smallskip}
\hline
\end{tabular}

Notes to Table 4.

$(a), (b), (c), (e), (g) $ and $(i)$ see the comments on the
radio properties of the radio galaxies in the A3528 complex 
in Section 4.1.

$(d)$ R = 3.03 for this identification, however we consider it
reliable since the radio emission is extended (see Table 3) and the
optical counterpart falls within the radio isodensity contours.

$^{(f)}$ This identification is unclear. We tentatively classified the radio 
source as a WAT.  See comments on the extended background 
sources in the A3528 region in Section 4.2.

$^{(h)}$ On the basis of the 13 cm image we classify this radio source
as a background FRII. The two optical counterparts given here are
located in the vicinity of the assumed core (see Section 4.2).

\end{flushleft}
\end{table*}
%
% END TABLE 4
%____________________________________

% FIGURE 4

\section{The radio emission in the A3528 region}

Most of the Shapley radio galaxies (6/9)
belong to A3528, two  are located in A3532, and the remaining
one is located in a peripheral region between A3530 and A3532.
It is noteworthy that we detect no radio emission from
the core of A3530, despite the presence of optically peculiar
galaxies in its centre.
We also note the lack of radio
sources between the centres of A3530 and A3532. Given the good
spatial coverage of our survey, this can not be attributed to
a decrease in sensitivity in this region.
From Figure \ref{fig:sources} and 
Figure \ref{fig:wedge} it is clear that most radio galaxies
in the A3528 complex are located in the densest optical regions.

All radio galaxies in the A3528 chain have logP$_{22cm}$ (W Hz$^{-1}) \ge$ 22,
and the radio power for two of them,
namely J1254$-$2913 in A3528S and J1257$-$3021 in A3532, is close to
the transition between FRI and FRII radio galaxies (Fanaroff \& Riley 1974).

\subsection{Extended radio galaxies in the A3528 complex}

\begin{figure*}
\epsfbox{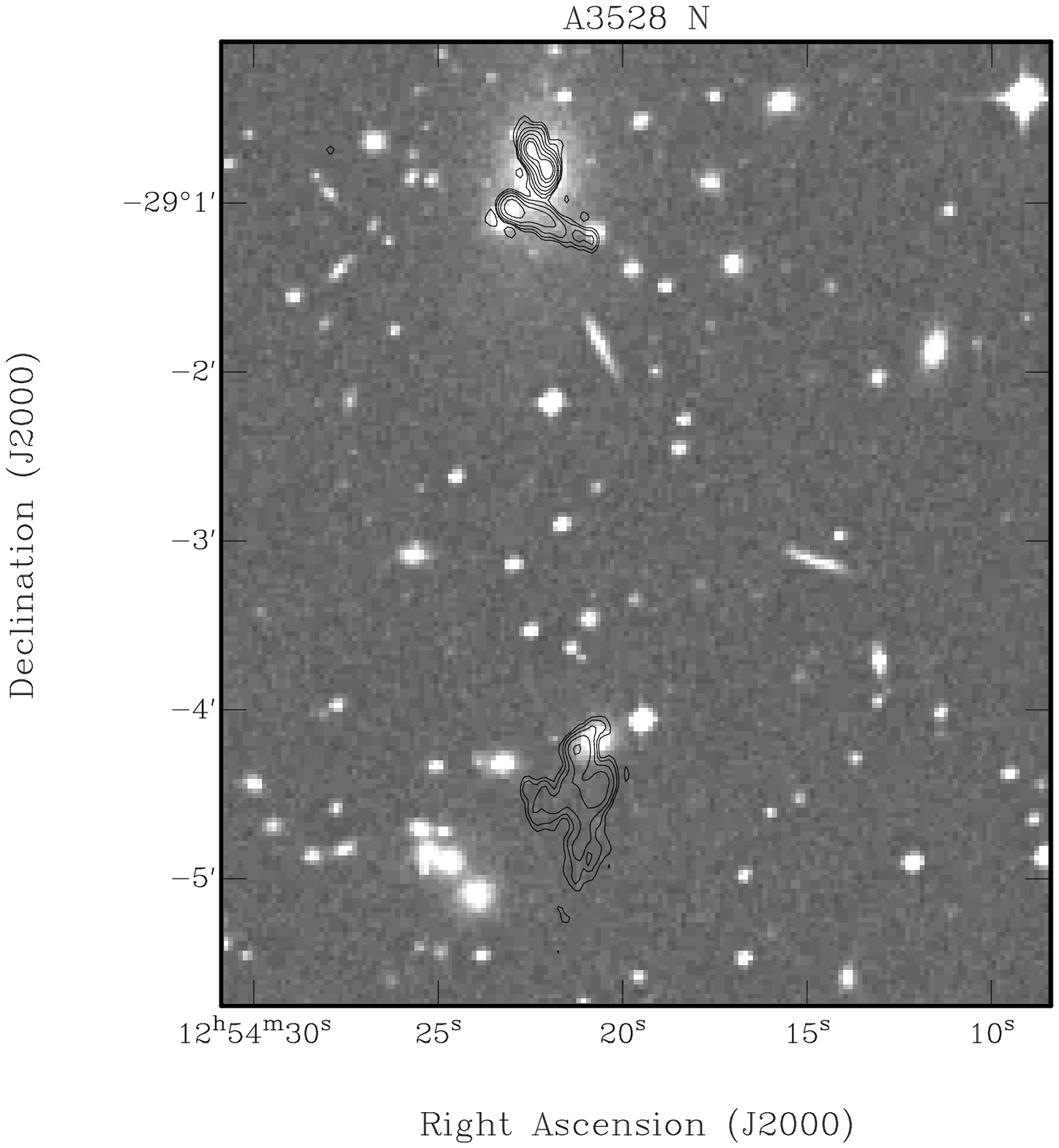}
\caption[]{13 cm image of the central part of A3528N overlaid
on the optical DSS image. The radio galaxies shown are
(from top to bottom): J1254$-$2900, J1254$-$2901a, J1254$-$2904.
Contours are -0.5, 0.5, 1, 2, 4, 8, 16 mJy b$^{-1}$. The
restoring beam is 5.73$\times$3.62, p.a. $1.99^{\circ}$.}
\label{fig:a3528n}
\end{figure*}

\begin{figure*}
\epsfbox{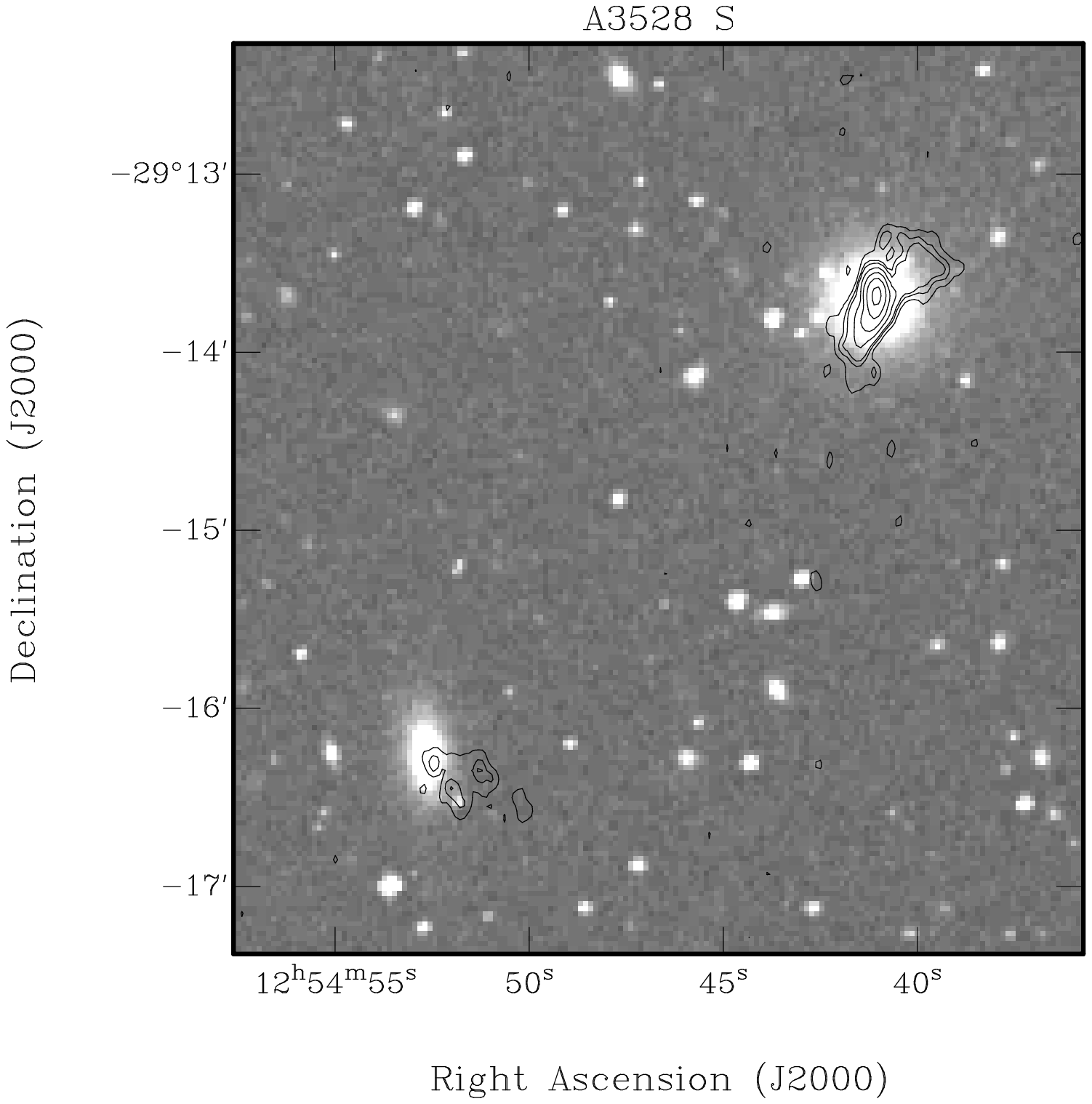}
\caption[]{13 cm image of the central part of A3528S overlaid
on the optical DSS image. The radio galaxies shown are
J1254$-$2913 and  J1254$-$2916.
Contours are -0.35, 0.35, 0.7, 1, 2, 4, 8, 16 mJy b$^{-1}$.
The restoring beam is $6.59^{\prime\prime}\times3.58^{\prime\prime}$, 
p.a. $2.28^{\circ}$.}
\label{fig:a3528s}
\end{figure*}

In Figures \ref{fig:a3528n} and \ref{fig:a3528s} we show the 13 cm
images of the central regions of A3528N and A3528S respectively,
overlaid on the DSS optical image.
Three extended radio galaxies are located
in the northern component A3528N, namely J1254$-$2904, J1254$-$2900 and
J1254$-$2901a, and two more are located in A3528S,
i.e. J1254$-$2913 and J1254$-$2916.
While all radio galaxies in A3528N are very close in velocity space,
the two in A3528S are separated by $\sim$ 2500 km s$^{-1}$
and possibly belong to two different groups in A3528S (see Table 4).

\noindent
Another extended radio galaxy, J1257$-$3021, is associated with one
of the two nuclei in the dumb-bell system dominating A3532.
 
Here we will briefly summarise the most relevant physical properties of 
these six extended radio galaxies on the basis of our 22 cm and 13 cm 
observations and then compare them with the properties of the
intracluster gas as derived from the X-ray observations.

% FIGURE 5

% FIGURE 6

\subsubsection{Morphologies and radio properties}

\noindent
J1254$-$2900 (Fig. \ref{fig:a3528n}) is associated with the dominant 
cD galaxy
in A3528N and is completely embedded within the envelope of 
the optical galaxy.
RHP98 classified this source as double,
and high resolution and high frequency observations carried out
with the VLA and ATCA (Venturi et al., in preparation)
confirm that it is a mini-FRII radio source, with
a radio nucleus and two jets culminating in hot spots,
though its total power falls within the range of FRI radio galaxies.
The source has a moderately steep spectrum over the frequency
range of  our observations, i.e. $\alpha^{13}_{22} = 0.9$.

\medskip
\noindent
J1254$-$2901a was classified as a head-tail radio galaxy by RHP98,
who also postulated that the radio source may have been 
triggered by the galaxy's passage close to the cluster centre.
Inspection of the superposition between the radio image and the
Digitised Sky Survey (Fig. \ref{fig:a3528n})
reveals that the radio peak is displaced from the centre of the optical
counterpart. This is most likely a resolution effect, i.e. the nucleus
of the radio emission is not coincident with the peak of our 22 cm and
13 cm images, but it could also be  
interpreted as an extreme effect of ram
pressure exerted on the radio emission due to the galaxy motion
through the very dense medium in the centre of A3528N.
The source is extended $\sim 60$ kpc east-west.
The total spectral index of this source, computed in the
range of frequencies presented in this paper, is 
$\alpha^{13}_{22} = 1.2$. We point out that it should be considered
an upper limit, since the lack of short spacings at 13 cm is likely
to lead to an underestimate of the extended flux density.
Furthermore the incomplete separation of this source from J1254$-$2900
at 22 cm adds uncertainties in the 22 cm flux density.

\medskip
\noindent
J1254$-$2904 (Fig. \ref{fig:a3528n}) 
is a very interesting radio galaxy in A3528N.
The source is $\sim$ 4 arcmin away from the cluster centre, in the 
region where the shock front between the two merging systems A3528N
and A3528S is expected to be located (S96).
The 13 cm image shows an extension in the direction of the galaxy 
\#2063  (see also Figure 3 in RHP98). 
On the assumption that this is the host galaxy we classify
the radio source as a narrow-angle tail. 
This hypothesis is confirmed by observations at
higher frequency and resolution (Venturi et al., in preparation),
where the radio core is clearly detected. 
We therefore confirm
the reliability of the tentative optical identification
proposed by RHP98.
The source is extended $\sim$ 1.5 arcmin (i.e. $\sim 65$ kpc). 
Its spectrum is steep in the range
13$~-~$22 cm, i.e. $\alpha^{13}_{22} = 1.3$, and flattens at lower
frequencies. If we consider the VLA 92 cm flux density 
(S96) and the MOST 36 cm data (RHP98)
we derive $\alpha^{36}_{92} = 0.5$ and $\alpha^{22}_{36} = 1.1$.
We stress that the spectral index $\alpha^{13}_{22}$ should be 
considered only an upper limit to its true
value, since its extended morphology suggest
that we may have lost extended flux at 13 cm due to the
poor short spacing u-v coverage of the present observations. 

\medskip 
\noindent
J1254$-$2913 (Fig. \ref{fig:a3528s})
is the second most powerful radio galaxy in the
A3528 complex and it is associated with the dominant cD galaxy
in A3528S. The radio source is dominated by a strong compact component
surrounded by extended emission whose shape is reminiscent 
of wide-angle tail sources (see also RHP98). Its
size, $\sim 35 \times 20$ kpc,
barely exceeds the optical extent of the associated galaxy.
Using the data at other wavelengths available 
in the literature (RHP98 and S96) we derive a spectral index 
$\sim 1.0$, constant within the errors,
in the range 92$~-~$13 cm.
% The equipartition parameters for this source are similar to those found
% J1254$-$2900, at the centre of A3528N, i.e. 
% B$_{eq} \sim 4.3~~ \mu$, U$_{eq} \sim 1.7\times10^{-12}$ erg cm$^{-3}$ and
% P$_{eq} \sim 1.1\times10^{-12}$ dyn cm$^{-2}$.

\medskip
\noindent
J1254$-$2916 is the weakest extended radio galaxy in the A3528 complex.
The higher resolution 13 cm image (Fig. \ref{fig:a3528s}),
together with the 92 cm VLA image from S96,
suggests that the source is a head-tail, with a faint
compact component coincident with the associated cluster galaxy.
Higher frequency and resolution observations (Venturi et al.
in preparation) confirm that this is the nucleus of the radio galaxy.
From our flux density measurements we derived a total flat 
spectrum with $\alpha_{13}^{22} = 0.41$.
The source has low surface brightness and
diffuse morphology in both bands presented in this paper,
and comparison with the 20 cm total flux density obtained from
inspection of the NVSS suggests that we may have lost extended
flux both at 22 cm and 13 cm. For this reason we believe that
the derived spectral index should be taken with care.
We point out that for the other extended
sources we are presenting in this section, the NVSS 20 cm and the ATCA
22 cm flux density measurements are in good agreement
(see Section 3.2). 
Given the uncertainty in the spectral index value, we defer any
estimate of the physical parameters in J1254$-$2916 to a future
paper.

\begin{figure}
\epsfysize=6cm
\epsfxsize=\hsize
\epsfbox{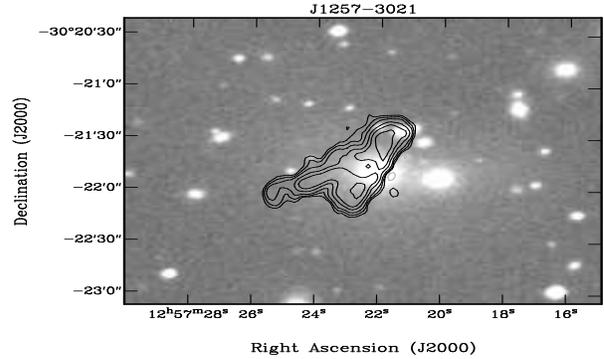}
\caption[]{13 cm radio image of J1257$-$3021 in A3532 superimposed 
on the DSS optical frame. Contour levels are:
-0.5, 0.5, 1, 2, 5, 10, 20 mJy b$^{-1}$.
The restoring beam is $5.7^{\prime\prime} \times 3.5^{\prime\prime}$
in p.a. 1$^{\circ}$.}
\label{fig:j1257-3021}
\end{figure}

\medskip
\noindent
J1257$-$3021 is the most powerful radio galaxy in the A3528 
complex, and is associated with the brighter nucleus of the
dumbell galaxy located at the centre of A3532 
(see Figure \ref{fig:j1257-3021}).
Its radio morphology is intermediate between FRI and FRII radio galaxies,
consistent with its 22 cm total radio power. 
The peculiar asymmetric shape of the tails suggests that
it could be a wide-angle tail source with its emission in a plane
closely aligned with the line of sight to the observer. 
The core of the radio emission is undetectable both at 22 cm and in the
full resolution 13 cm image. We used the 6 cm flux density value
for this source given in Gregorini et al. (1994), and computed 
the spectral index over the range 6$~-~$22 cm, obtaining 
$\alpha \sim 0.85$, typical for this type of radio source.

\subsubsection{Comparison with the properties of the intracluster gas}

\medskip
In order to derive estimates of the physical conditions
in the relativistic plasma for the radio galaxies presented
in this section, we computed 
the equipartition values of the magnetic field B$_{eq}$,
energy density U$_{eq}$ and internal non-thermal pressure P$_{eq}$
assuming a cylindrical geometry, a filling factor $\Phi$=1
and a ratio between protons and electrons K=1. The values
we computed are given in Table 5. We point out that these are
global values, and refer to the whole source. 
We did not include J1254$-$2916 because of the major uncertainties in the
spectral index value. For all the remaining extended sources
we adopted the value $\alpha_{22}^{13}$ given above for each source.
The dependance of B$_{eq}$ and P$_{eq}$ on the spectral index is not
critical, so even though the values we derived for 
$\alpha_{22}^{13}$ should be considered upper limits, the estimates
of the intrinsic physical parameters are not affected
significantly.

% TABLE 5.
%\setcounter{table}{5}
\begin{table}
\caption[]{Properties of the extended Shapley galaxies}
\begin{flushleft}
\begin{tabular}{lccc}
\hline\noalign{\smallskip}
 Source       & B$_{eq}$    & P$_{eq}$ & P$_{gas}$ \\
              & 10$^{-6}$ G & 10$^{-11}$ dyn cm$^{-2}$ & 
10$^{-11}$ dyn cm$^{-2}$ \\  

\hline\noalign{\smallskip}
J1254$-$2900  & 4.5  & 0.13  &  5.2 \\
J1254$-$2901a & 2.7  & 0.047 &  4.5 \\
J1254$-$2904  & 1.6  & 0.016 &  1.0 \\ 
J1254$-$2913  & 4.3  & 0.11  &  7.5 \\
J1257$-$3021  & 4.5  & 0.13  &   *  \\
\noalign{\smallskip}
\hline
\end{tabular}

\end{flushleft}
\end{table}
%------- end of Table 5

\medskip
Our estimates for the equipartition parameters 
for all radio galaxies given in Table 5  
fall within the range of values
found for a sample of tailed radio galaxies in clusters 
(Feretti, Perola \& Fanti 1992). We note that 
in the case of J1254$-$2904
the values we found are amongst the lowest. 

\noindent
For comparison with the equipartition parameters,
in Table 5 we also report the thermal pressure of the intracluster
gas at the projected distance of the radio galaxies with respect
to the cluster centre, derived from Henriksen \& Jones (1996) and
scaled for our choice of cosmological 
parameters.
None of the radio galaxies is in pressure equilibrium with the
external thermal gas. 
For J1254$-$2900 and J1254$-$2913 projection effects are 
expected to be negligible, since they are both associated
with the dominant galaxies in A3528N and A3528S respectively,
so we believe that the derived unbalance is real.
The case for J1254$-$2904 is most dramatic even accounting for
projection effects, since P$_{eq}$ and P$_{gas}$ differ by 
almost two orders 
of magnitude. Assuming the
gas distribution given by Henriksen \& Forman (1996), J1254$-$2904
would be in pressure equilibrium with the external gas at $\sim 32$
arcmin from the cluster centre, i.e. $\sim$ 1.3 Mpc.
No information is available in the literature for the temperature
and gas density distribution in A3532.

A more detailed analysis will be possible with XMM observation
already awarded for this cluster complex.

% figure 7

\medskip
\subsection{Extended background sources in the A3528 region}

Among the extended background sources revealed in the A3528 field,
three exhibit a peculiar morphology. In this section we
describe them and present their global properties.

\subsubsection{J1254$-$3042 and J1255$-$2934}

\begin{figure}
\epsfysize=5cm
\epsfxsize=\hsize
\epsfbox{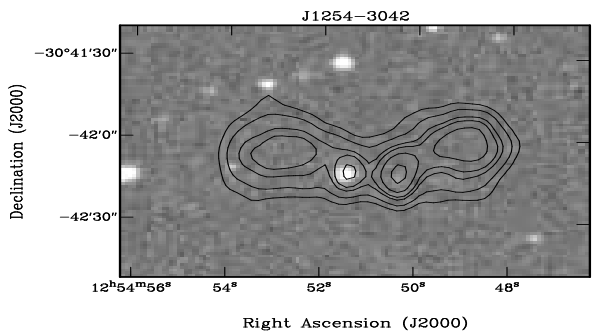}
\caption[]{22 cm radio image of J1254$-$3042 superimposed on the DSS
optical frame. The contours are -0.35, 0.35, 0.75, 1.5, 2, 3, 5, 
10 mJy b$^{-1}$.
The restoring beam is $10.9^{\prime\prime} \times 6.3^{\prime\prime}$,
in p.a. $-1.5^{\circ}$.}
\label{fig:j1254-3042}
\end{figure}

\begin{figure}
\epsfysize=6cm
\epsfxsize=\hsize
\epsfbox{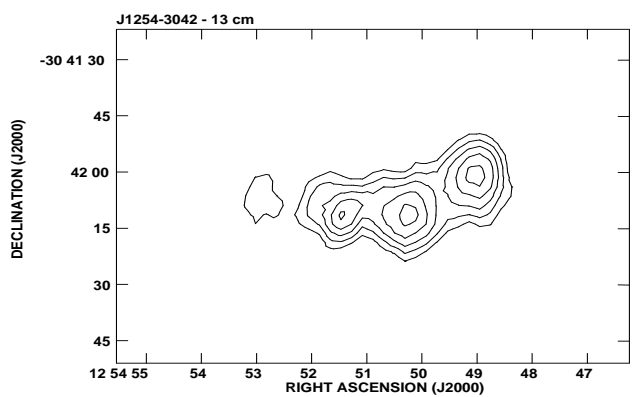}
\caption[]{13 cm natural weighted radio image of J1254$-$3042. 
The image is
not corrected for the primary beam attenuation.
Contour levels are: -0.3, 0.3, 0.4, 0.5, 0.6, 0.7 mJy b$^{-1}$.
The restoring beam is $18.6^{\prime\prime} \times 11.0^{\prime\prime}$
in p.a. $4.4^{\circ}$.}
\label{fig:j1254_13cm}
\end{figure}

The 22 cm image of the radio source J1254$-$3042, 
shown in Figure \ref{fig:j1254-3042}, is suggestive of a 
double-double type morphology (Lara et al. 1999) in the shape of a 
wide-angle tail, however the
optical identification is unclear, since two 
very faint objects coincide with the two inner peaks of
the radio emission (see Table 4). 
From the  13 cm natural weighted image shown in Figure \ref{fig:j1254_13cm},
it is clear that the source is very asymmetric in this band, with 
the eastern lobe much fainter than the western one. 

\begin{figure}
\epsfysize=5cm
\epsfxsize=\hsize
\epsfbox{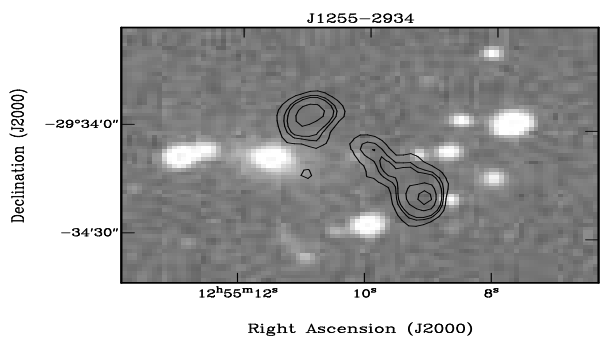}
\caption[]{13 cm radio image of J1255$-$2934 superimposed on the DSS
optical frame. The contours are -0.35, 0.35, 0.75, 1, 1.5, 3, 5, 10, 15, 17 
mJy b$^{-1}$.
The restoring beam is $6.6^{\prime\prime} \times 3.7^{\prime\prime}$,
in p.a. 2.3$^{\circ}$.}
\label{fig:j1255-2934}
\end{figure}

J1255$-$2934 has a typical FRII morphology, with a one-sided jet
and two hotspots with similar flux density. The full resolution
13 cm image is given in Figure \ref{fig:j1255-2934}.
A very faint optical counterpart (see Table 4) is visible, close to 
the central radio peak between the two hot spots,
but the identification
with this object is uncertain. It is possible that the counterpart
responsible for the radio emission is an absorbed high redshift
galaxy.

% figure 8

% figure 9

% figure 10

\medskip
\subsubsection{The extended unidentified radio source J1256$-$3009}
\begin{figure}
\epsfysize=5cm
\epsfxsize=\hsize
\epsfbox{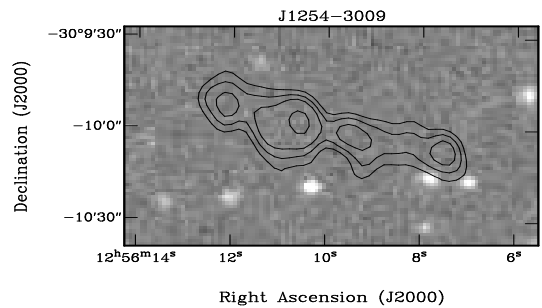}
\caption[]{22 cm radio image of J1256$-$3009 superimposed on the DSS
optical frame. The contours are -0.3, 0.3, 0.5, 0.75, 1, 1.5 mJy b$^{-1}$.
The restoring beam is $10.9^{\prime\prime} \times 6.3^{\prime\prime}$,
in p.a. $-1.5^{\circ}$.}
\label{fig:unident}
\end{figure}

The 22 cm radio density contours of J1256$-$3009 are shown in Figure 
\ref{fig:unident}, superimposed on the DSS optical frame.

% figure 11

This radio source has a linear morphology,
with an extension of $\sim$ 1.5 arcmin. It is located north 
of the axis connecting the centres of A3530 and A3532, at a distance
of $\sim 14^{\prime}$ from the centre of A3530 and $\sim 19^{\prime}$
from the centre of A3532. The DSS optical image shows the
presence of a few faint optical objects in the vicinity of the radio emission,
but none of them is obviously associated with the radio source.
Unfortunately the source was not detected at 13 cm, possibly because
of its low surface brightness combined with  
lower sensitivity at its distance from the field centre 
(field \#3), therefore no hint on its nature can be derived from
the spectral index information. The source could be 
a double-double
powerful radio galaxy associated with a distant object fainter than 
the DSS limit.
Another possibility is that J1256$-$3009 is diffuse cluster emission within 
the Shapley Concentration. On  this hypothesis the source would
be $\sim$ 60 kpc in extent, with a power logP$_{22cm}$ (W Hz$^{-1}$) = 22.87.
We point out that J1256$-$3009 lies in the
region where a merging shock due to the interaction between A3530 and
A3532 is expected. We plan to extend our  study of this source in order to 
disentangle its nature.

%____________________________________

\section{Statistical Investigations}

\subsection{Radio source counts}

Because of the primary beam attenuation, the sensitivity of
our observations is a function of the distance from the
centre of each field. For this reason our survey is not
complete at a flux density limit of 0.8 mJy.
In order to carry out our statistical investigations,
we restricted our study to all radio sources within a radius of 
17.5 arcmin from each field centre, with flux density 
S$_{22cm}\ge$ 1.6 mJy.
At such distance, the primary beam attenuation implies a flux
density correction of a factor of 2, so 
sources with uncorrected S$_{22cm}$ = 0.8 mJy have a true flux density of 
1.6 mJy. 
This complete sample contains 58 radio sources detected within
an area of 1.04 deg$^2$. 

\begin{figure}
\epsfysize=5cm
\epsfxsize=\hsize
\epsfbox{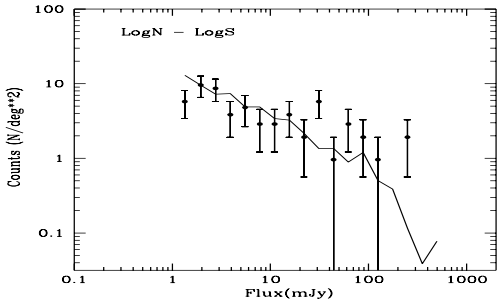}
\caption[]{LogN$-$LogS. Dots refer to counts in the A3528 complex; the
solid line represents the distribution of the background
(Prandoni et al. 1999)}
\label{fig:logn}
\end{figure}

The LogN-LogS for the A3528 complex, computed in the flux
density range 1$-$1440 mJy, is shown in
Figure \ref{fig:logn}. The errors are poissonian.
The solid line in Figure \ref{fig:logn} represents the normalized
LogN-LogS for the background (Prandoni et al. 1999)
taken from a 22 cm ATCA survey, covering an area of
25.82 deg$^2$ and counting a total of 1752 radio
sources above the flux density limit S$_{22cm}$ = 1 mJy.

Inspection of Figure \ref{fig:logn} suggests that the two
distributions have similar shapes,
and we conclude that the number of radio sources in the
A3528 complex is in good agreement with the background radio
source counts. We found 54 sources above a flux density
of 2 mJy, consistent with the 48 expected from the
background counts.
%We applied a Kolmogoroff-Smirnov (KS) test and
%found that the probability for A3528 complex radio source
%counts to be a fluctuation of the background is p=0.736.

We note that the optical bidimensional overdensity of the A3528 complex,
with respect to the background in a nearby region, is $\sim$ 2.5.
This means that if the radio emission probability per galaxy remained 
constant in the Shapley Concentration with respect to the field, 
we would expect to detect a significant excess at $\sim 5\sigma$ above
the background.

We conclude that the radio source counts in the A3528 complex
are dominated by the background counts.
This result is similar to that obtained for the
A3558 complex (Paper III), the other major
cluster merger in the core of the Shapley Concentration. 

% FIGURE 12

\subsection{Radio luminosity function}

The influence of the environment on the radio
properties of galaxies should reflect on the radio luminosity
function (RLF), changing either its shape or the normalisation.
In order to test if the merger in the A3528 complex
affects the probability of a galaxy to develop a radio
source, we computed the RLF for the early-type
galaxies in this region, for comparison with the RLF
computed by LO96 
for a complete sample of Abell clusters, and with that
derived for the A3558 complex
in the Shapley Concentration (Paper III).

All nine radio galaxies in this region satisfy the criteria
adopted in Paper III for a straightforward comparison with LO96,
i.e. optical counterpart brighter than B$_J = -18.48$
and S$_{22cm} \ge$ 2.2 mJy (see Paper III for details). Therefore, we
could use all of them in the computation of the RLF. 
At the distance of the A3528 complex S$_{22cm} = 2.2$ mJy corresponds
to logP$_{22cm}$(W Hz$^{-1}$) = 21.78.

\noindent
For the galactic absorption we adopted a mean value of 0.3,
obtained from the maps of Burnstein \& Heiles (1984). The 
corresponding magnitude limit for the A3528 complex is
therefore  b$_J$ = 17.70.

\noindent
Inspection of Table 4 shows that two radio sources with S$_{22cm} > 2.2$
mJy are associated with galaxies with  b$_J < $ 17.40 but 
without redshift information. In particular,
J1254$-$2858b is associated with a b$_J$ = 17.33 magnitude
galaxy and J1254$-$3019 with a b$_J$ = 14.08 object. Given their
optical magnitudes and using the spectroscopic data available in
this region, we determined the probability that
these two galaxies belong to the Shapley Concentration
as 70\% and 66\% respectively. We therefore consider them
Shapley galaxies and include them in the computation
of RLF. Assuming they are located at the average redshift of
the A3528 complex, i.e. v = 16000 km s$^{-1}$ (see Section 2),
we derived  power of logP$_{22cm}$ (W Hz$^{-1}$) = 21.98 and 22.24
for J1254$-$2858b and J1254$-$3019 respectively.
To summarise, we used 11 radio galaxies for the computation of RLF.

For the normalisation of the RLF we 
counted all galaxies brighter than b$_J$ = 17.70 and 
13000 $\le cz \le$ 19000 km s$^{-1}$ within a radius of  32 arcmin
from each pointing centre, and covered the whole region of our
radio observations (see Figure \ref{fig:point}). 
There are two sources of errors in the proper number of galaxies. 

\noindent {\it (a)} 
Of the 124 candidate galaxies in our surveyed region, only
114 have a measured redshift. Of these, the number with a measured
velocity in the range of the Shapley Concentration is 89.
Assuming that the fraction of Shapley
galaxies remains the same for the remaining 10 galaxies
without redshift, we conclude that there are 97 galaxies.

\noindent
{\it (b)} 
Another problem concerns the fraction of 
early-type galaxies, since a morphological classification
is uncertain at this distance.
A galaxy-type classification based on the optical spectra 
carried out on a subsample of galaxies in the A3528 complex
(Baldi et al., in preparation), shows that only $\sim$ half of them
are early type. 

\begin{figure}
\epsfysize=5cm
\epsfxsize=\hsize
\epsfbox{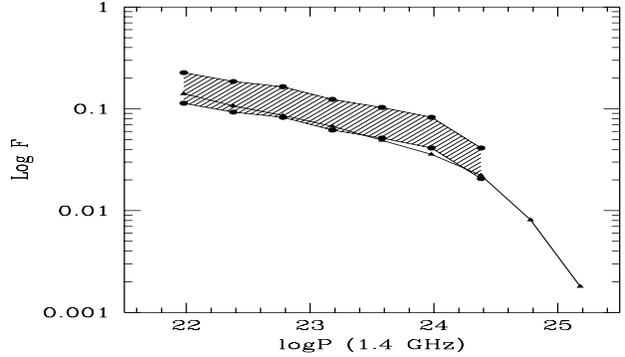}
\caption[]{Integral radio luminosity function.
Filled circles represent the two extremes of the RLF 
derived for the A3528 complex and
the shaded area represents the region where the real RLF must lie.
Filled triangles give the RLF derived by LO96 for Abell clusters.}
\label{fig:fdl}
\end{figure}

\noindent
For these reasons we computed 
two RLFs, the first assuming that all 97 galaxies are early type,
and the second assuming that only a fraction of 50\%, i.e. 49, 
is early type. We are aware these are two extreme cases. 
However, it seems reasonable that the true RLF lies within these two 
limits.
The results of our analysis are given in Table 6, where the 
fractional and integral RLF for the two cases is reported.
The RLF is plotted in Figure \ref{fig:fdl}, where the shaded region represents 
the area where the true RLF is located. The lower envelope
represent the case where all galaxies are early type, while the
upper envelope corresponds to the case of 50\% early-type galaxies.

% TABLE 6.
%\setcounter{table}{1}
\begin{table*}
\caption[]{Fractional and integral RLF of A3528 cluster complex}
\begin{flushleft}
\begin{tabular}{ccccc}
\hline\noalign{\smallskip}
 $\Delta$ logP$_{22}$ & Fractional RLF & Integral RLF & Fractional RLF & 
Integral RLF \\
                      & (no late type) & (no late type) & (50\% late type) &
(50\% late type)      \\
%\noalign{\smallskip}
\hline\noalign{\smallskip}
 21.78$-$22.18 & $2/97$  & 0.113 & 2/49 & 0.224 \\
 22.18$-$22.58 & $1/97$  & 0.093 & 1/49 & 0.184 \\
 22.58$-$22.98 & $2/97$  & 0.082 & 2/49 & 0.163 \\
 22.98$-$23.38 & $1/97$  & 0.062 & 1/49 & 0.122 \\
 23.38$-$23.78 & $1/97$  & 0.052 & 1/49 & 0.102 \\
 23.78$-$24.18 & $2/97$  & 0.041 & 2/49 & 0.082 \\
 24.18$-$24.58 & $2/97$  & 0.021 & 2/49 & 0.041 \\
\noalign{\smallskip}
\hline
\end{tabular}

\end{flushleft}
\end{table*}
%------- end of Table 6

For comparison, the luminosity function derived by LO96 for Abell 
clusters is also given in Figure \ref{fig:fdl} (triangles).
The two RLFs are in good agreement 
in the power range 21.78$\le$ logP $\le$ 24.58,
both in shape and normalisation.
We note a lack of radio galaxies with  
logP$_{22}$ (W Hz$^{-1})>$ 24.58 in the A3528 complex. 

% FIGURE 13

\section{Discussion and Conclusions}
This work is part of a larger ongoing project, whose main
purpose is to study the effects of cluster mergers on the 
statistical properties of radio galaxies, and on the
formation of cluster-type radio sources, such as radio
halos and relics, which are believed to be
related to major merger events. The target of our
study is the central region of the Shapley Concentration,
the largest supercluster in the nearby Universe, where
spectroscopic and X-ray observations lend support to 
earlier suggestions that cluster mergers play a major role (see
for instance Raychaudhury 1991, Zucca et al. 1993, Ettori
et al. 1997, Bardelli et al. 1998, Bardelli et al. 2000).

The results of our 22 cm ATCA survey in the A3528 complex
can be summarised as follows:

\medskip
\noindent 
{\it (a)} The radio properties of the three clusters in
the chain differ markedly from one other.
A3528, considered to be a pre-merging cluster,
is very active at radio wavelengths, with
five extended and distorted radio galaxies located at the
centres of its two subcondensations. Conversely, we detected
no radio sources from A3530 or from the region between
the cores of A3530 and A3532, where a merger shock is 
expected.

\medskip
\noindent
{\it (b)} The radio source counts are consistent, within the
uncertainties, with the
background counts down to the mJy level, implying that the 
radio source density does not scale with the optical density, 
which is a factor of 2.5 higher in the A3528 complex compared to 
the local background. This result is similar to what we 
obtained for the merging cluster complex A3558 (Paper III),
and confirms our earlier conclusions that the radio source 
counts do not  trace the underlying optical density. Therefore, very
high density regions would not be spotted on the basis
of radio surveys.

\medskip
\noindent
{\it (c)} The radio luminosity function for early-type galaxies
is consistent, within the errors, with the RLF derived by LO96
for a large sample of cluster ellipticals. This suggests
that the dense and unrelaxed environment of the A3528 region has no
influence on the probability of an elliptical galaxy
to become a radio source. 

It is interesting to compare this result with the RLF
obtained for the A3558 merging complex, which is
significantly lower than LO96 (Paper III).
We propose that the different shapes and normalisations
of the RLF in these two complexes of merging clusters are 
due to the different stages of the merger itself.

On the basis of the gas temperature gradients within A3528,
S96 first proposed that A3528 is an early merger, where
A3528N and A3528S have not yet undergone a major core-to-core
encounter. More
recently Bardelli et al. (2000) reinforced this assessment
from a bi- and tri-dimensional analysis of clusters
and groups in this region, emphasising that the central parts
of the individual galaxy clusters in the chain are not far from 
virialization, and further suggesting that 
the effects of the merger are not yet evident on the galaxy
population and cluster dynamics.
The good agreement between the RLF derived in this paper for the
A3528 cluster region and that for early type galaxies in Abell
clusters could be interpreted in the light of
Bardelli et al. (2000), i.e. the early stage of the merger
has not yet affected the radio properties of cluster galaxies,
which behave remarkably similarly to ellipticals in other environments. 

The situation in the A3558 region is quite different, since 
the observational evidence suggests that the cluster complex
has already undergone the first core-to-core encounter 
for two massive clusters, and further accretion of small
groups is at an advanced stage (Bardelli et al. 1998b). The 
excess of blue galaxies in the position where a shock front
in the merging structure is expected, and the RLF lower than
LO96, could be interpreted as the 
effects of the advanced merger, which has already 
had time to affect the galaxy optical and 
radio emission properties.
One implication of this interpretation is that a cluster merger 
ultimately anticorrelates
with the probability of a galaxy developing a radio source,
either by switching off previously existing radio galaxies or
inhibiting the formation of radio-loud AGNs.

%We conclude that 
%the results of the present work, coupled with previous work
%in the literature, suggest that the statistics are still too
%low to clearly understand the role of cluster mergers
%on the radio emission from galaxies in clusters. We are presently
%working on a radio survey other sets of merging clusters 
%in order to increase the information
%available.

\vskip 1.0truecm
\noindent
{\bf Aknowledgments}

We warmly thank Dr. Isabella Prandoni for insightful
discussions and for providing data before publication.

\noindent
T.V. acknowledges the receipt of the CNR-CSIRO grant Prot. N. 088864.
The Australia Telescope Compact Array is operated by the CSIRO 
Australia Telescope National Facility.

\noindent
The Australia Telescope Compact Array is operated by the
Australia Telescope National Facility.

{}

\end{document}